\documentstyle[epsfig]{mn}

\newcommand{\sigx}{\sigma_{\tilde\epsilon_1}}
\newcommand{\sigy}{\sigma_{\tilde\epsilon_2}}
\newcommand{\sige}{\sigma_{\epsilon}}
\newcommand{\gred}{\tilde g_{\rm r}}
\newcommand{\teps}{\tilde\epsilon}
\newcommand{\wave}{\langle w\rangle_z}
\newcommand{\wsqr}{\langle w^2\rangle_z}

\title[Galaxy-galaxy lensing and cluster lens reconstruction]
{A simultaneous maximum likelihood approach for galaxy-galaxy
lensing and cluster lens reconstruction}

\author[Bernhard Geiger and Peter Schneider]
       {Bernhard Geiger$^{1,2}$ and Peter Schneider$^1$\\
        $^1$Max-Planck-Institut f\"ur Astrophysik,
        Karl-Schwarzschild-Stra{\ss}e 1,
	85740 Garching bei M\"unchen, Germany\\
	$^2$Institut d'Astrophysique de Paris,
	98bis Boulevard Arago, 75014 Paris, France}

\date{Submitted 1998 April 21}
\pagerange{\pageref{firstpage}--\pageref{lastpage}}
\pubyear{1998}

\begin{document}

\label{firstpage}

\maketitle

\begin{abstract}
In a previous paper we investigated means for constraining the mass
distribution of cluster galaxies by weak lensing. We concluded
that a comprehensive method should treat the lensing effects of
individual cluster galaxies and those resulting from a general cluster
component simultaneously. To this end we now develop a
non-parametric maximum likelihood cluster reconstruction algorithm
that can implicitly take into account the presence of cluster galaxies. 
The method includes an entropy-like regularization prescription and
directly uses the ellipticities of individual source galaxy images 
as observables rather than relying on an averaged ellipticity
field. The mass distribution of cluster galaxies is described by 
parametrized models. For each set of galaxy parameters the cluster
reconstruction algorithm allows to determine the best representation  
of the global underlying cluster component that is consistent with the
presence of the cluster galaxies and the observed image ellipticities
of background galaxies. Tests with simulations yielded convincing and
robust results. 

We applied the method to a WFPC2 image of the cluster Cl0939+4713 and
obtained a detection of the lensing effects of luminous elliptical
cluster galaxies. We consider this application as a successful test of
our technique. However, the small size of the image we analysed
does not yet allow to draw strong conclusions regarding the mass
distribution of cluster galaxies.   
\end{abstract}


\section{introduction}
Weak gravitational lensing was suggested by Tyson et~al. (1984) as a
means to obtain information on the mass distribution of galaxies at large
radial distances. The first detection of weak distortions of the
images of background galaxies generated by the tidal gravitational
field of intervening galaxies was reported by Brainerd, Blandford \& Smail
(1996). Griffiths et~al. (1996) studied the galaxy-galaxy
lensing effects in the {\it Medium Deep Survey}\/, and Dell'Antonio \&
Tyson (1996) and Hudson et~al. (1997) analysed the {\it Hubble Deep Field}\/. 
Natarajan et~al. (1997) detected the weak lensing effects induced by
cluster galaxies in a mosaic of HST-images of the cluster AC114.

In this paper we continue our discussion of methods to constrain the
mass distribution of cluster galaxies from weak lensing presented in
Geiger \& Schneider (1998, hereafter referred to as Paper~I). For the
maximum likelihood method described in that paper we first obtained a
reconstruction of the cluster mass distribution using `conventional'
techniques that rely on the inversion of an integral equation. Then we
added parametrized models for the mass distribution of galaxies to the
cluster component and applied empirical mass subtraction prescriptions
to keep the total mass of the system constant. This approach proved to
be workable in the outskirts of clusters where the surface mass
density is low. In the non-linear lensing regime in the cluster
centre, however, the result turned out to be sensitive to the details
of the mass reconstruction and subtraction procedures. For values of
the galaxy model parameters that imply a very large fraction of the total mass
attributed to galaxies, conceptional problems arise regardless of the
(non-)linearity of the lensing effects or the distance from the
cluster centre.   

In order to overcome these problems we decided to extend the maximum likelihood
approach to the problem of determining the mass distribution of the cluster
component. This allows to include prior information on the presence of
the cluster galaxies directly into the cluster reconstruction procedure.  
We follow the work of Bartelmann et~al. (1996) and use the two-dimensional 
gravitational potential as the fundamental physical quantity describing the
properties of the cluster lens. Likelihood methods for cluster
reconstruction from weak lensing were also described by Squires \& 
Kaiser (1996) and Bridle et~al. (1998). These methods use a grid of averaged 
image ellipticities of background galaxies as observational
input. However, the averaging procedure already destroys the
information on galaxy scales that we are interested in and therefore
we are obliged to use the ellipticities of individual source galaxy
images directly as observables. Such an approach for cluster
reconstruction was investigated in detail by Seitz, Schneider \&
Bartelmann (1998), and some elements of the technical implementation
of our method are similar to their study. 

In Section~\ref{meth} we describe the combined method for cluster lens
reconstruction and galaxy-galaxy lensing that we have developed.
The reliability of the method is confirmed by an application to
simulated observations presented in Section~\ref{appsim}. In
Section~\ref{appreal} we use our techniques to analyse a WFPC2 image
of the cluster Cl0939+4713, which yields a detection of the lensing
effects induced by elliptical cluster galaxies. Section~\ref{disc}
concludes the paper with a discussion of the results. 

\section{method}\label{meth}
We continue to use the concepts and definitions introduced in
Paper~I. In general the symbols for the surface mass density, the
shear, and the lensing potential should be understood as referring to
(hypothetical) sources located at infinite redshift. For simplicity we
omit the $\infty$-subscript for the relevant quantities in this paper
if additional subscripts are present.

\subsection{Description of the cluster component}
The lensing properties of the cluster component are described by the 
dimensionless scalar potential $\psi_{\infty}$. The surface mass density 
$\kappa_{\rm C}$ and the shear 
$\gamma_{\rm C}=\gamma_{1\rm C}+{\rm i}\,\gamma_{2{\rm C}}$
are combinations of second derivatives of this potential:
\[
\kappa_{\rm C}=\frac{1}{2}(\psi_{,11}+\psi_{,22}), \ \
\gamma_{1\rm C}=\frac{1}{2}(\psi_{,11}-\psi_{,22}), \ \
\gamma_{2\rm C}=\psi_{,12}\;.
\]
In practice, we specify the potential $\psi_{\alpha\beta}$ on a 
$(n+2)\times(n+2)$ grid ($\alpha,\,\beta=0,\dots,n+1$), and calculate the 
cluster contribution to the surface mass density 
$\kappa_{{\rm C}\,\alpha\beta}$ and the shear 
$\gamma_{{\rm C}\,\alpha\beta}=\gamma_{1{\rm C}\,\alpha\beta}+
{\rm i}\,\gamma_{2{\rm C}\,\alpha\beta}$ on a $n\times n$ grid 
($\alpha,\,\beta=1,\dots,n$), using discrete versions of the second
derivatives:
\[
\kappa_{{\rm C}\,\alpha\beta}=\frac{1}{2}
(\psi_{\alpha+1\,\beta}+\psi_{\alpha-1\,\beta}+
\psi_{\alpha\,\beta+1}+\psi_{\alpha\,\beta-1}-4\,\psi_{\alpha\beta})\;,
\]
\[
\gamma_{1{\rm C}\,\alpha\beta}=\frac{1}{2}
(\psi_{\alpha+1\,\beta}+\psi_{\alpha-1\,\beta}-
\psi_{\alpha\,\beta+1}-\psi_{\alpha\,\beta-1})\;, \ \ {\rm and}
\]
\[
\gamma_{2{\rm C}\,\alpha\beta}=\frac{1}{4}
(\psi_{\alpha+1\,\beta+1}+\psi_{\alpha-1\,\beta-1}-
\psi_{\alpha+1\,\beta-1}-\psi_{\alpha-1\,\beta+1})\;.
\]
The index values $\alpha,\beta=1$ and $\alpha,\beta=n$ correspond to opposite 
corners of a square field of view. The lensing parameters 
$\kappa_{i{\rm C}}$ and $\gamma_{i{\rm C}}$ at the positions of individual 
background galaxy images $i$ are obtained by bilinear interpolation of their 
values on the adjacent grid points. They can therefore be expressed as linear 
combinations of the potential components:
\begin{equation}
\kappa_{i{\rm C}}=\sum_{\gamma,\delta=0}^{n+1}\,
a_{i\gamma\delta}\,\psi_{\gamma\delta}\ \ \ \
\mbox{and}\ \ \ \ \gamma_{i{\rm C}}=\sum_{\gamma,\delta=0}^{n+1}\,
b_{i\gamma\delta}\,\psi_{\gamma\delta}\;.
\end{equation}
For each background galaxy image the coefficients $a_{i\gamma\delta}$ and 
$b_{i\gamma\delta}$ are non-zero only for 12, respectively 16, neighbouring 
grid points. 

The total surface mass density $\kappa_i$ and shear $\gamma_i$ are calculated 
by adding the contributions $\kappa_{i{\rm G}j}$ and $\gamma_{i{\rm G}j}$ 
from the cluster galaxies ${\rm G}_j$ to those of the global cluster component:
\begin{equation}
\kappa_i=\kappa_{i{\rm C}}+\sum_{j=1}^{N}\kappa_{i{\rm G}j}\ \ \ \
\mbox{and}\ \ \ \
\gamma_i=\gamma_{i{\rm C}}+\sum_{j=1}^{N}\gamma_{i{\rm G}j}\;.
\end{equation}
The values $\kappa_{i{\rm G}j}$ and $\gamma_{i{\rm G}j}$ depend on the 
relative positions of background images and cluster galaxies and, of course,
on the mass distribution of the cluster galaxies. As in Paper~I we use a 
truncated singular isothermal sphere as a model for their dark matter 
distribution and apply the scaling laws
\begin{equation}\label{scaling}
\sigma=\sigma_{\star}\,\left(\frac{L}{L_{\star}}\right)^{1/\eta}\ \ \ \
\mbox{and}\ \ \ \ s=s_{\star}\,\left(\frac{L}{L_{\star}}\right)^{\nu}
\end{equation}
with $\eta=4$ and $\nu=0.5$ to relate the model parameters to those of
an $L_{\star}$-galaxy with velocity dispersion $\sigma_{\star}$ and
cut-off radius $s_{\star}$. This choice of parametrization was
critically discussed in Paper~I and in practice different strategies
could be more appropriate.   
However, the principles of the method described here do not depend on the 
details of the galaxy parametrization scheme, and for the rest of this section 
$\sigma_{\star}$ and $s_{\star}$ can be more generally interpreted as 
representing a given set of parameters used to describe the mass distribution 
of the cluster galaxies.

The likelihood function is defined as the product of the probability densities
for the observed image ellipticities of the background galaxies:
\begin{equation}
{\cal L}(\sigma_{\star},s_{\star},\,\{\psi_{\alpha\beta}\})=
\prod_i\,p_{\epsilon}(\epsilon_i\,|\,\kappa_{i},\,\gamma_{i})\;.
\end{equation}
It depends on the galaxy parameters $\sigma_{\star}$ and $s_{\star}$ 
and on the description of the global cluster component provided by the
potential values $\psi_{\alpha\beta}$ on the grid points. 

\subsection{Probability density distributions}\label{spdd}
In Paper I we calculated and illustrated the probability density distribution
$p_{\epsilon}(\epsilon\,|\,\kappa_{\infty},\,\gamma_{\infty})$ for the image 
ellipticities of gravitationally lensed background galaxies. For the present 
method it turned out to be necessary to approximate $p_{\epsilon}$ by a 
simple analytical function for numerical reasons. 

In the case of a single redshift plane for the background galaxies, or if an
estimate for the redshift of individual galaxies is available (e.g.,
photometric redshifts), the expectation value for image ellipticities is
determined by the reduced shear: 
\begin{equation}
\langle\epsilon\rangle_{\epsilon_{\rm s}}=\hat g:=
\left\{
\begin{array}{lll}
g & {\rm for} & |g|\leq1 \\
1/g^{\star} & {\rm for} & |g|>1\;.
\end{array}
\right.
\end{equation}
Contours of the probability density distribution in the ellipticity parameter 
space $\epsilon$ are fairly circular (see Fig.~5 of Paper~I) and we show in 
Appendix~\ref{appprob} that the dispersion of the distribution in `shear 
direction' is equal to the dispersion perpendicular to this direction in the 
ellipticity coordinates. Therefore, the Gaussian distribution
\begin{equation}\label{pappsimp}
p_{\epsilon}(\epsilon\,|\,g)\approx\frac{1}{\pi\,\sige^2(\hat g)}\,
{\rm e}^{\displaystyle -\frac{|\epsilon-\hat g|^2}{\sige^2(\hat g)}}
\end{equation}
represents a convenient approximation to the probability density distribution
which reproduces its first and second moments. The dispersion $\sige(\hat g)$
as a function of $\hat g$ can be approximated by a simple analytical fit to
numerical results similar to the discussion further below.

\begin{figure}
    \epsfxsize=84mm
    \epsffile{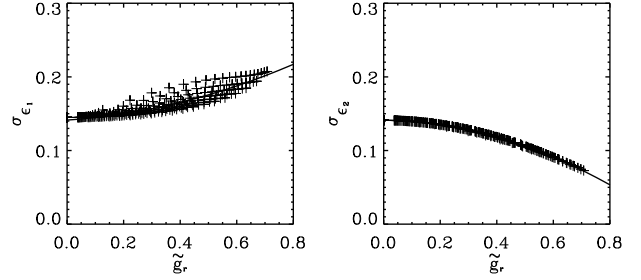}
\caption{Approximations for the dispersion of the probability density
$p_{\epsilon}(\epsilon)$. The dispersions $\sigx$ and $\sigy$ were calculated 
for a grid of parameter values of the quantities $\kappa_{\infty}$ and 
$|\gamma_{\infty}|$. The crosses in the plots show the results as a function of
$\gred$ for all data points with $|g_{\infty}|<1$. 
The solid lines show fits to these data points using quadratic functions.
The results displayed in this figure were calculated for the intrinsic
ellipticity ($\sigma_{\epsilon_s}=0.2$) and redshift distributions that were 
used in the simulations of Paper~I.}
\label{gred}
\end{figure}

In the case of a redshift distribution for the source galaxies it is more
difficult to specify simple expressions to approximate the probability density
distributions. In this paper we restrict the application of our method to 
images located in non-critical regions of clusters. In this case we can use 
the approximation (Seitz \&~Schneider 1997) 
\[
\langle\epsilon\rangle_{\epsilon_{{\rm s},z}}\approx
\frac{\wave\,\gamma_{\infty}}{1-\frac{\wsqr}{\wave}\,\kappa_{\infty}}
\]
for the expectation value of the image ellipticities. In the presence of a 
source redshift distribution the probability distribution for the image 
ellipticities is elongated in the `shear direction' (see Fig.~5 of Paper~I), 
i.e., the dispersion in this direction is larger than in the perpendicular 
direction. Hence we model the probability distribution with a Gaussian of the 
form
\begin{eqnarray}\label{papp}
p_{\epsilon}(\epsilon\,|\,\kappa_{\infty},\,\gamma_{\infty}) & \approx &
\frac{1}{2\pi\,\sigx(\gred)\,\sigy(\gred)}\;\cdot \nonumber \\
 & & \cdot\;
{\rm e}^{-\frac{1}{2}\displaystyle \frac{(\teps_1-\gred)^2}{\sigx^2(\gred)}}\,
{\rm e}^{-\frac{1}{2}\displaystyle \frac{\teps_2^2}{\sigy^2(\gred)}}\;,
\end{eqnarray}
where $\gred$ is defined as
\[
\gred:=
\frac{\wave\,|\gamma_{\infty}|}{1-\frac{\wsqr}{\wave}\,\kappa_{\infty}}\;,
\ \ \ {\rm and}\ \ \
\teps=\teps_1+{\rm i}\,\teps_2:=
\frac{\gamma^{\star}_{\infty}}{|\gamma_{\infty}|}\,\epsilon
\]
expresses the image ellipticities in `local shear coordinates', i.e., in a 
coordinate system in which the imaginary part of the (reduced) shear vanishes.
We approximate the dependence of the dispersions $\sigx$ and $\sigy$ on $\gred$
by quadratic functions:
\[
\sigx(\gred)\approx\frac{1}{\sqrt{2}}\,\sigma_{\epsilon_{\rm s}}+
c_{11}\,\gred+c_{12}\,\gred^2\;,\ \ \ {\rm and}\ \ \
\]
\[
\sigy(\gred)\approx\frac{1}{\sqrt{2}}\,\sigma_{\epsilon_{\rm s}}+
c_{21}\,\gred+c_{22}\,\gred^2\;.
\]
For a given intrinsic ellipticity and redshift distribution of the sources
the constants are determined by a fit to numerical results as shown in
Fig.~\ref{gred}.

\subsection{Likelihood maximization}
The logarithm of the likelihood function is given by
\begin{equation}
l(\sigma_{\star},s_{\star},\,\{\psi_{\alpha\beta}\}):=\ln\,{\cal L}=
\sum_i\,\ln\,p_{\epsilon}(\epsilon_i\,|\,\kappa_{i},\,\gamma_{i})\;.
\end{equation}
For a given set of cluster galaxy parameters $(\sigma_{\star},s_{\star})$
the likelihood is maximized with respect to the potential values 
$\psi_{\alpha\beta}$ on the grid points in order to determine the best 
representation of the global cluster component that is consistent with the
presence of the cluster galaxies as specified by the model. 
Repeating this maximization procedure for varying galaxy parameters allows to
determine the likelihood $l(\sigma_{\star},s_{\star})$ as a function of the 
galaxy model and to derive confidence regions in the parameter space.

Specifying the parameters for the galaxy mass model determines the mass 
$M_{\rm G}$ contained in galaxies in absolute numbers:
\begin{equation}
M_{\rm G}(\sigma_{\star},s_{\star})
=\sum_{j=1}^{N}M_{{\rm G}j}(L_j,\,\sigma_{\star},s_{\star})\;.
\end{equation}
Here $L_j$ denotes the cluster galaxy luminosity, but it can be more generally
interpreted as representing a set of galaxy characteristics that are used to
specify the mass model for individual galaxies. On the other hand, the total 
mass $M$ of the galaxy cluster, or equivalently the average surface mass 
density $\bar\kappa_{\infty}$, is provided by a mass reconstruction. The 
likelihood function defined above only uses information on the shapes of 
background images and therefore it is invariant under the mass sheet 
transformation (see Paper I). However, we assume that this degeneracy can be 
broken by other means and so the total cluster mass within the field of view 
is fixed. In order to include the information on the total mass into the 
maximization procedure we adopt the following strategy: The fraction of mass 
contained in the global cluster component and not associated with galaxies is 
denoted as
$f_{\rm C}(\sigma_{\star},s_{\star}):=1-M_{\rm G}(\sigma_{\star},s_{\star})/M$
and depends on the galaxy model parameters. Multiplication of the likelihood 
function with the factor
\begin{equation}
p_{\bar\kappa_{\rm C}}(\bar\kappa_{\rm C})=
\frac{1}{\sigma_{\bar\kappa_{\rm C}}\,\sqrt{2\pi}}\,{\rm e}^{-\frac{1}{2}
\displaystyle \frac{(\bar\kappa_{\rm C}-f_{\rm C}\,\bar\kappa_{\infty})^2}
{\sigma_{\bar\kappa_{\rm C}}^2}}
\end{equation}
ensures that the average surface mass density
\[
\bar\kappa_{\rm C}=\frac{1}{n^2}
\sum_{\alpha,\beta=1}^n\kappa_{{\rm C}\alpha\beta}
\]
of the reconstructed global cluster component reproduces the correct mass
fraction $f_{\rm C}(\sigma_{\star},s_{\star})$. This corresponds to
adding the term 
\begin{equation}
l_{\bar\kappa_{\rm C}}=\ln p_{\bar\kappa_{\rm C}}=-
\frac{(\bar\kappa_{\rm C}-f_{\rm C}\,\bar\kappa_{\infty})^2}
{2\,\sigma_{\bar\kappa_{\rm C}}^2}+{\rm constant}
\end{equation}
to the logarithm of the likelihood. The value of the parameter 
$\sigma_{\bar\kappa_{\rm C}}$ is only of numerical interest and has been 
chosen as $0.001$ for the applications described later in this paper.

The lensing quantities, surface mass density and shear, are second derivatives
of the potential $\psi_{\infty}$, and therefore adding linear
functions to the potential does not change the mass distribution and
the likelihood. It turned out that the presence of these invariance
transformations does not cause practical problems in our
implementation of the method. In principle they can be suppressed
during the maximization procedure, for example by fixing the potential
values at three corners of the field of view as in Bartelmann et al. (1996).

In order to prevent the cluster component from fitting noise and from 
exhibiting structures on small scales we also add an entropy-like 
regularization term of the form                                     
\begin{equation}
S=\sum_{\alpha,\beta=1}^n\ln\hat\kappa_{{\rm C}\alpha\beta}
\end{equation}
(see e.g. Narayan \& Nityananda 1986) with
\[
\hat\kappa_{\rm C\alpha\beta}=\tilde\kappa_{\rm C\alpha\beta}\;/
\sum_{\gamma,\delta=1}^n\tilde\kappa_{{\rm C}\gamma\delta}\ \ \ \
\mbox{and}\ \ \ \
\tilde\kappa_{\rm C\alpha\beta}=
\kappa_{\rm C\alpha\beta}/\kappa_{\rm P\alpha\beta}
\]
to the logarithm of the likelihood function. An additional benefit of
this regularization term is that it ensures the reconstructed
surface mass density to be positive. The factor $\kappa_{\rm
P\alpha\beta}$ allows to include prior information on the mass
distribution into the regularization. 

A conjugate gradient algorithm from Press et al. (1992) is then used in order 
to maximize the quantity
\begin{equation}\label{lhat}
\hat l(\sigma_{\star},s_{\star},\,\{\psi_{\alpha\beta}\}):=
l+l_{\bar\kappa_{\rm C}}+\lambda\,S\;.
\end{equation}
This algorithm makes use of the derivatives 
$\partial\hat l/\partial\psi_{\alpha\beta}$ which are straightforward to 
compute analytically (see Appendix \ref{deriv}). The choice of the 
regularization parameter $\lambda$ allows to adjust the result between a pure
likelihood maximization and a reproduction of the prior mass distribution. 
In general we use the mass distribution from a 
conventional reconstruction as the prior. In addition, we calculate the
potential $\psi_{\rm P\alpha\beta}$ corresponding to this conventional 
reconstruction and use 
$f_{\rm C}(\sigma_{\star},s_{\star})\,\psi_{\rm P\alpha\beta}$ as the start
value for the maximization algorithm. This guarantees the correct mass 
fraction for the start value of the cluster component as a function of the 
galaxy model parameters and therefore reduces the computational effort for the
maximization procedure. 

\section{Application to simulations}\label{appsim}
In Paper~I we divided the data of our simulations into two subsets according 
to the position of the background galaxies within the field of view: A central
sub-field containing the highly non-linear lensing regime, and the outskirts 
of the cluster where the surface mass density is comparatively low.
For computational reasons we restrict the application of the generalized
likelihood method presented in this paper to the central data region that 
covers $2\farcm5\times2\farcm5$ and includes roughly 90 cluster galaxies and
240 background galaxies. We use a conventional inversion method as explained in
Paper~I to reconstruct the mass distribution within the total field of view
($10'\times10'$) of the simulations.  
The lensing potential $\psi_{\infty}$ can be determined by integrating
the surface mass  density appropriately. We then apply the likelihood
procedure described in the previous section taking the surface mass
density and the potential in the central sub-field as regularization
prior and start value, respectively. The reconstructed mass
distribution also specifies the average $\bar\kappa_\infty=0.32$ of
the total surface mass density within the sub-field, which is required
for the term $l_{\bar\kappa_{\rm C}}$ ensuring the correct total mass. 

\begin{figure*}
    \epsfxsize=84mm
    \epsffile{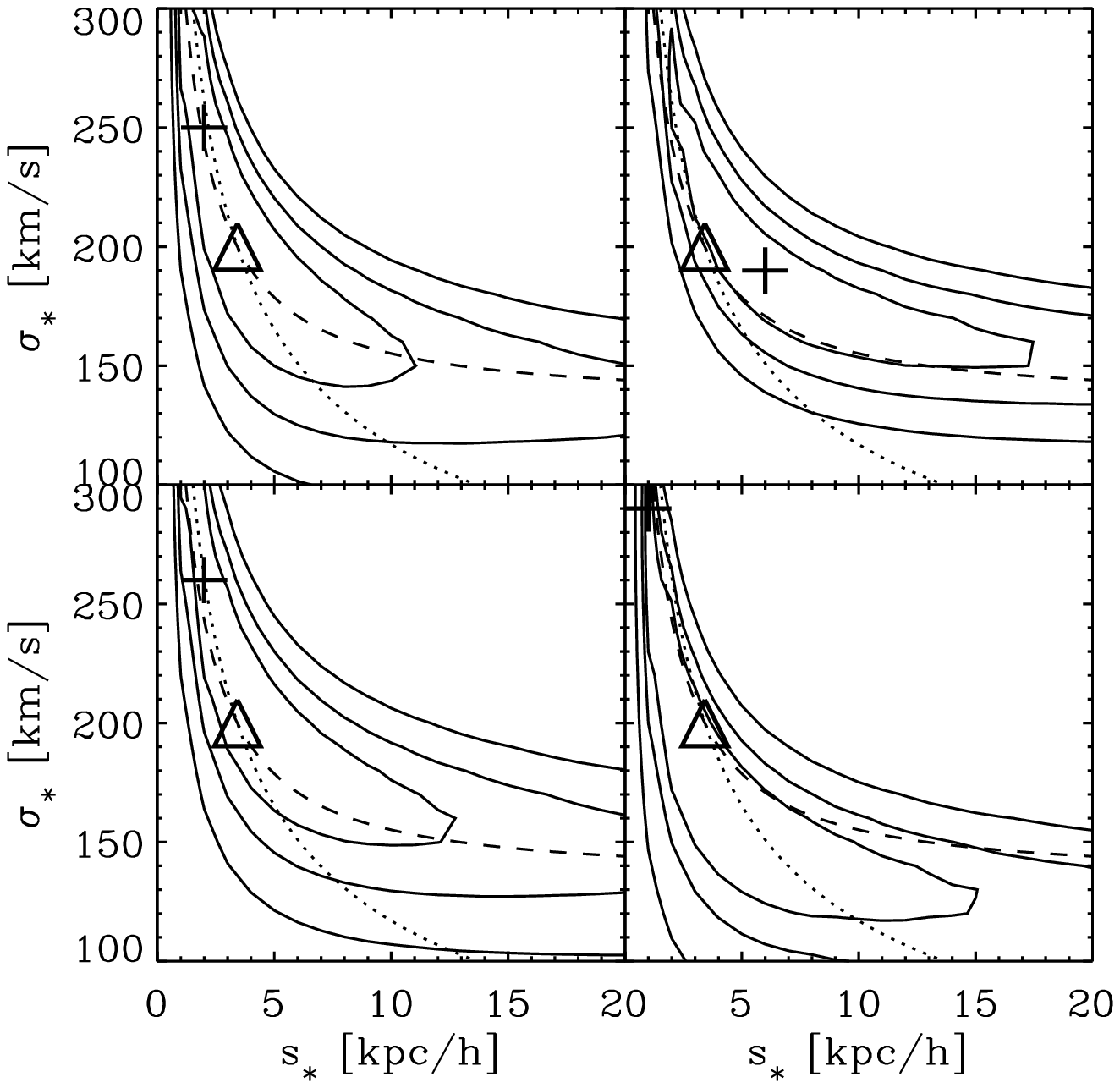}
    \epsfxsize=84mm
    \epsffile{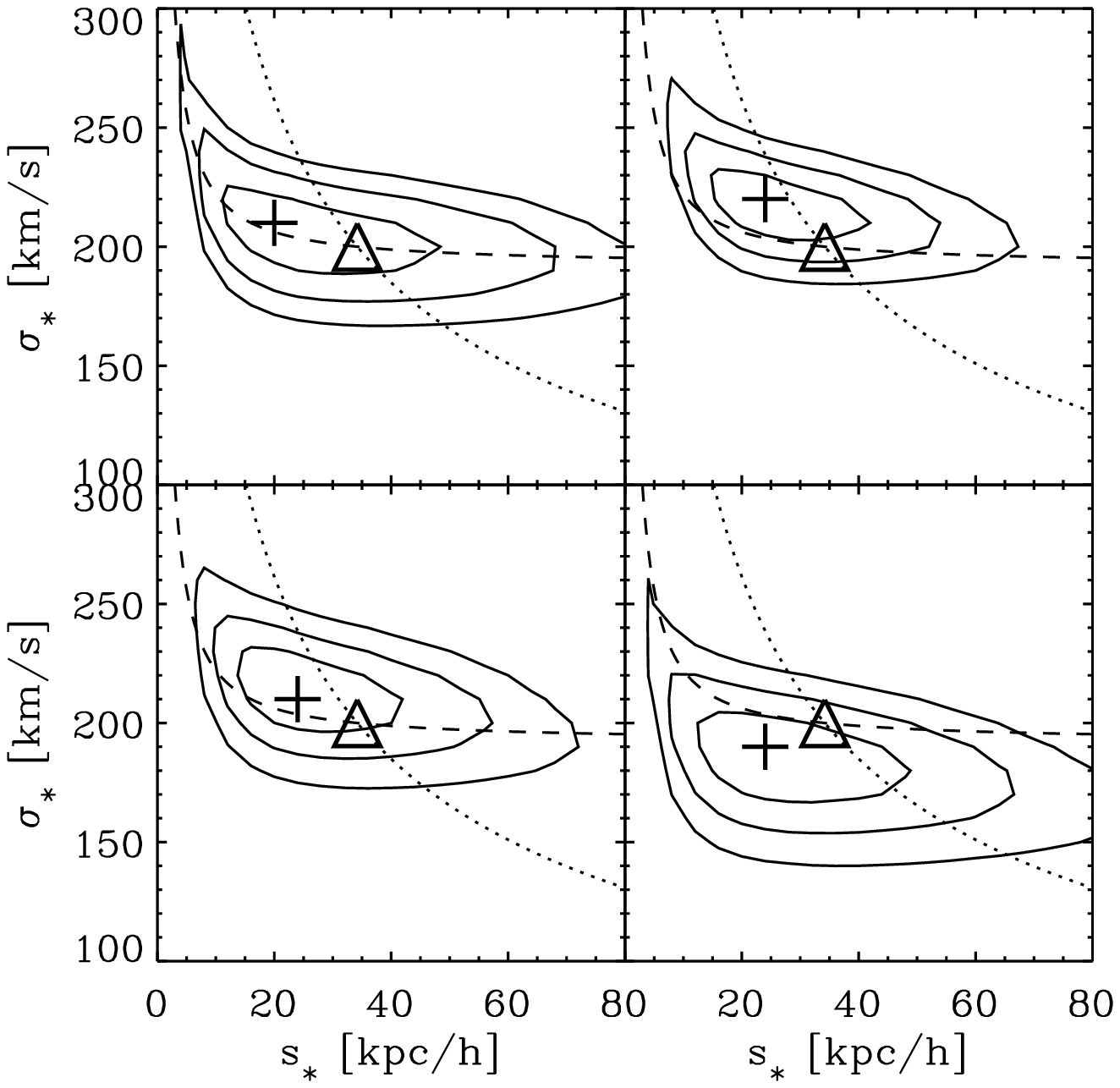}
\caption{Confidence regions for the velocity dispersion and the cut-off radius
calculated with the method described in this paper using a $5\times5$ grid and
the regularization parameter $\lambda=1$. The four different
realizations of simulated observational data are the same as in
Paper~I. This figure should be compared with the bottom sets of plots
in Fig.~7 of Paper~I. The plots on the {\bf left} are for the galaxy
input model with a cut-off radius of $s_{\star}=3.4h^{-1}\,{\rm kpc}$,
and the plots on the {\bf right} are for $s_{\star}=34h^{-1}\,{\rm kpc}$. 
The confidence contours are 99.73\%, 95.4\%, and 68.3\%, determined in
the way explained in Paper~I (without prior information). The triangle
denotes the input values and the cross marks the maximum of the
likelihood function. The dotted line connects models with equal total
mass, and along the dashed line the mass within a projected radius of
$5.4h^{-1}\,\rm{kpc}$ is constant.} 
\label{like3}
\end{figure*}

We explore the dependence of $l$ on the galaxy parameters, velocity dispersion
$\sigma_{\star}$ and cut-off radius $s_{\star}$, by maximizing the quantity 
$\hat l(\sigma_{\star},s_{\star})$ with respect to the potential values on the
grid points. Fig.~\ref{like3} shows confidence regions resulting from these
calculations for the same realizations of cluster and background galaxies as 
in Paper~I. The number of grid points used here is $5\times5$ which 
corresponds to a grid point separation of $68h^{-1}\,{\rm kpc}$. This provides
sufficient resolution for an adequate description of the global cluster mass 
distribution that had been used as an input for the simulations. The value of
the regularization parameter was chosen as $\lambda=1$. 

A comparison of the results for the input model with small galaxy 
haloes ($s_{\star}=3.4h^{-1}\,{\rm kpc}$) to those of Paper~I shows that the 
confidence regions for the galaxy parameters within the plotted range are very 
similar, but more extended with the new method. The simpler approach 
of Paper~I provides satisfying results when testing model parameters that 
imply a small fraction of the total mass in galaxies. For increasing galaxy 
mass fraction, however, the difference between the (appropriately scaled)
conventional mass reconstruction and the concept of a global underlying 
cluster mass component becomes larger. In the new method the cluster
component can adapt to the presence of potentially massive galaxies and
redistribute the matter in the cluster component accordingly. The additional
degrees of freedom lead to the larger extension of the confidence regions for 
increasing $s_{\star}$-values. (As in Paper~I the range of the cut-off
radius parameter was confined to small values in the plots in order to
emphasize the structure of the confidence regions close to the input
value, and because the inclusion of prior information on the velocity
dispersion as discussed in Paper~I could rule out large values for the
cut-off radius in this case.)

\begin{figure*}
    \epsfxsize=84mm
    \epsffile{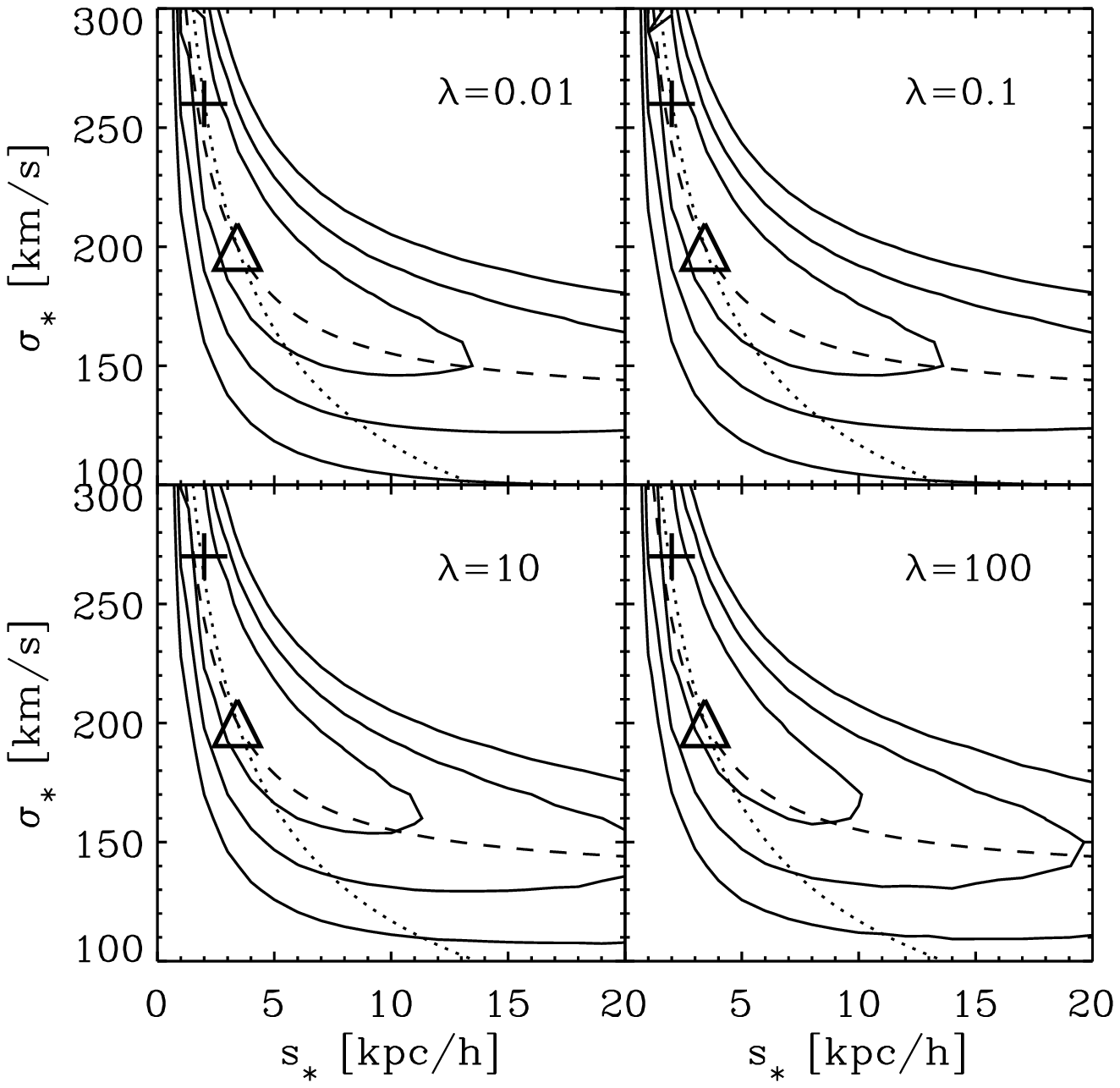}
    \epsfxsize=84mm
    \epsffile{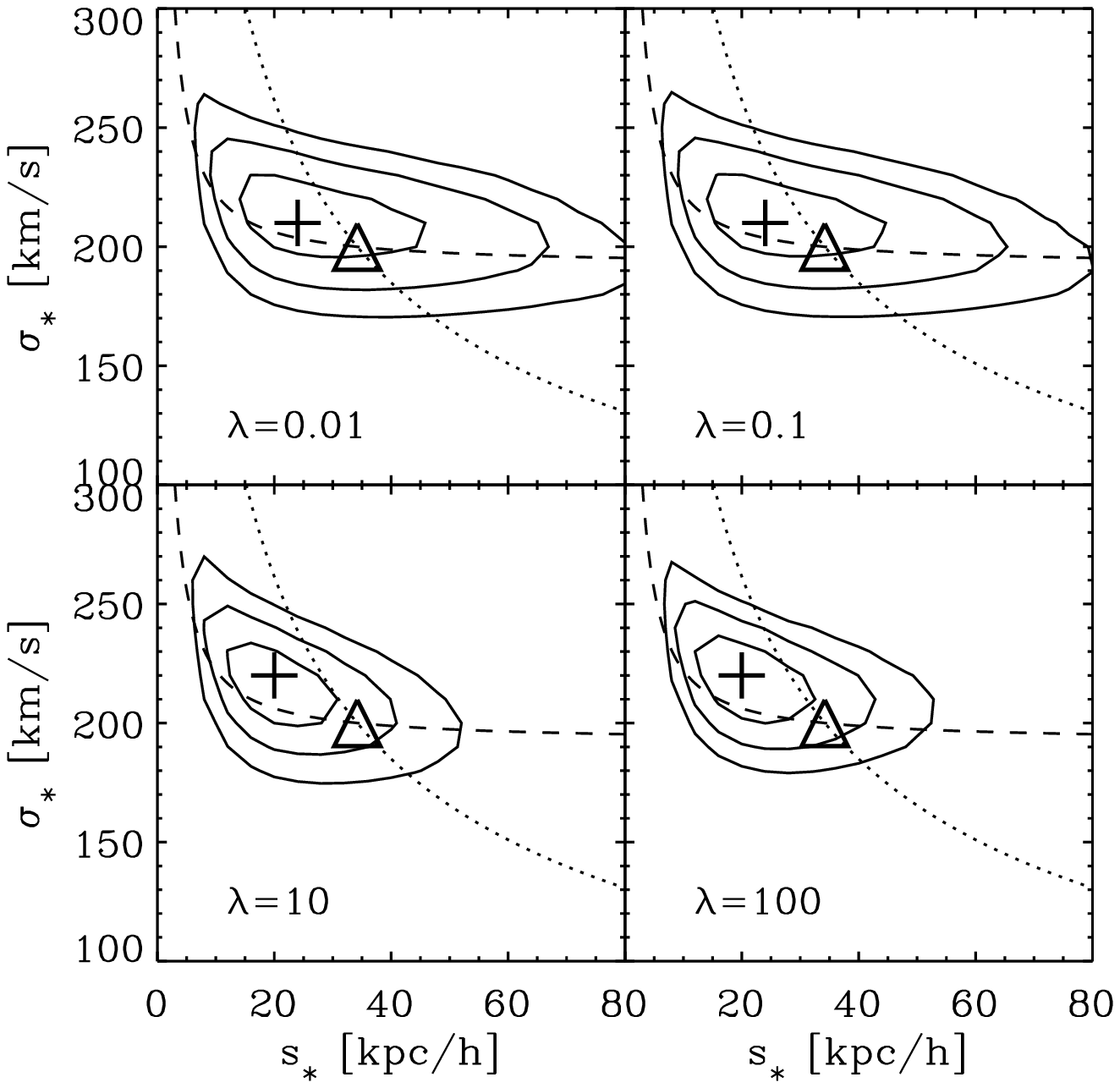}
\caption{Dependence of the confidence contours on the regularization
parameter $\lambda$. The random realization of cluster and background galaxies
is the same as for the bottom left plots in the panels of Fig.~\ref{like3}.
The values for $\lambda$ are 0.01 ({\bf top left}), 0.1 ({\bf top right}),
10 ({\bf bottom left}) and 100 ({\bf bottom right}).
The number of grid points is $5\times5$. The significance of the contours and 
the meaning of the symbols is analogous to Fig.~\ref{like3}.}
\label{like4}
\end{figure*}

In the model with $s_{\star}=34h^{-1}\,{\rm kpc}$ the galaxies contain a
considerable fraction of the total mass of the system. In Paper~I we 
investigated strategies for subtracting the galaxy masses from a conventional 
cluster reconstruction in order to use it as a description of the `global 
cluster component' in the likelihood method described there. For the highly
non-linear lensing regime in the cluster centre, however, this proved to be 
very difficult and the resulting confidence regions for the galaxy parameters 
turned out to be rather sensitive to the details of the subtraction procedure. 
These problems are nicely resolved by applying the generalized likelihood 
method presented in this paper, which takes the presence of the galaxies 
explicitly into account when determining the best representation of the global
cluster component for a given galaxy model. As discussed in Paper~I the
quantity that can be determined best with this lensing method is the mass
within the projected radius that corresponds to about the closest separation 
between cluster and background galaxies used in the analysis. As expected
the confidence regions resulting with the new method are extended
along lines of equal mass within this radius. Due to the freedom of
the cluster component to adapt to the galaxy model, the extension of
the confidence regions is larger than in the respective plots of Paper~I.

In Fig.~\ref{like4} we investigated the sensitivity of the results to the
choice of the regularization parameter. For large values of $\lambda$ the
`prejudice' that is provided by the conventional reconstruction and that
enters the method as the regularization prior, is highly weighted and
dominates over the likelihood term in equation
(\ref{lhat}). Regularizing too strongly therefore leads to the 
same kind of problems as discussed for the method of Paper~I, and the
confidence regions approach the solution of Paper~I for
$\lambda\rightarrow\infty$. This trend is more obvious for the input model 
with massive galaxies ($s_{\star}=34h^{-1}\,{\rm kpc}$), where the difference 
between the results of the two methods is larger. 
Reducing $\lambda$ enables the cluster component to adapt to changes in the
prescription for the galaxy model more liberally by maximizing the likelihood 
term $l$, and hence leads to a larger extension of the confidence regions for
the galaxy parameters. We discuss the question for the optimal choice of the 
regularization parameter below in context with the evolution of the absolute 
likelihood value. 

\begin{figure}
    \epsfxsize=84mm
    \epsffile{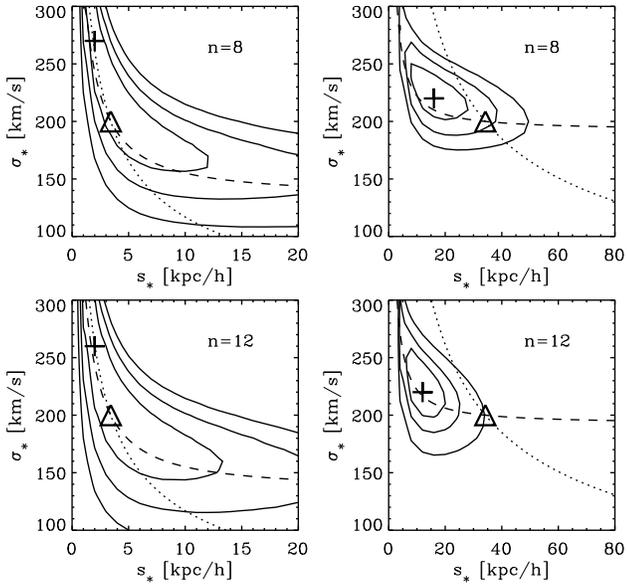}
\caption{Dependence of the confidence contours on the number of grid points.
The random realization of cluster and background galaxies is the same as for 
the bottom left plots of Fig.~\ref{like3}. The {\bf top} plots are for an 
$8\times8$ grid, and in the {\bf bottom} plots a $12\times12$ grid was used. 
The regularization parameter is $\lambda=1$. The significance of the contours 
and the meaning of the symbols is analogous to the previous figures.}
\label{like5}
\end{figure}

Fig.~\ref{like5} displays confidence contours calculated with a finer grid 
for describing the lensing potential of the cluster component. An $8\times8$ 
grid corresponds to a grid point separation of $39h^{-1}\,{\rm kpc}$, 
and a $12\times12$ grid gives $25h^{-1}\,{\rm kpc}$. In the case of the galaxy
model with small cut-off radius the result shown here is only slightly affected
by changing the grid, because the parameter range for the cut-off
radius covered by the plots still remains smaller than the grid point
separation. In the model with extended ($s_{\star}=34h^{-1}\,{\rm
kpc}$) dark matter haloes for the cluster galaxies, however, a refined
grid resolution already enables the cluster component to adapt within
a certain extent to the large haloes of luminous cluster galaxies. 
Models with a larger fraction of the mass in the cluster component and
smaller galaxy mass fraction have more freedom to adjust to
noise. This leads to a shift of the maximum likelihood
solution towards lower values for the $s_{\star}$-parameter and biases against
high cut-off radius values for which a large mass fraction is fixed in
galaxies and not available to fit noise. Solving this problem by applying
a stronger regularization to prevent the cluster component from fitting 
structures on `galaxy scales' is not straightforward because this increases the
weight given to the conventional reconstruction and leads to the same problems
as mentioned above. In order to reliably retrieve the input values of the 
simulations even with small grid point separations, it could be useful to 
explore strategies in which the prior is adapted during the maximization
process. However, it is a much more pragmatic point of view to accept these
difficulties as a manifestation of the fundamental conceptual problem of making
a clear-cut distinction between the galaxy mass distribution and a global
cluster component. In this philosophy the grid point distance provides the
scale for a technical separation of the galaxies from a more general mass 
component. 

\begin{figure}
    \epsfxsize=84mm
    \epsffile{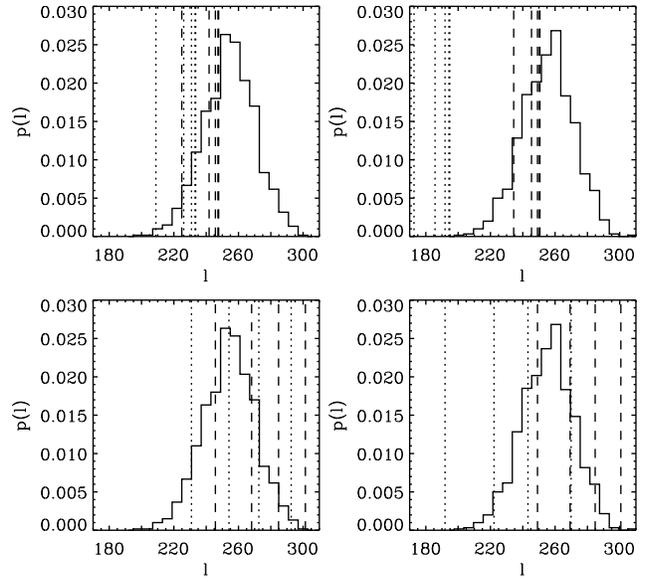}
\caption{Dependence of the maximum likelihood value on the regularization
parameter and on the number of grid points. The histograms show the
probability distribution of the maximum ${\rm Max}(l)$ of the
logarithm of the likelihood function calculated from the correct input
mass distribution and many different galaxy realizations. The plots on
the {\bf left} are for the galaxy input model with a cut-off radius of 
$s_{\star}=3.4h^{-1}\,{\rm kpc}$, and the plots on the {\bf right} are for
$s_{\star}=34h^{-1}\,{\rm kpc}$. Dashed vertical lines indicate the value of
$l$ at the maximum of the likelihood function, and dotted vertical lines
denote its value for a reconstruction without cluster galaxies. The {\bf
bottom} plots depict the dependence on the number of grid points for
$\lambda=1$. With increasing $l$ the lines represent a $5\times5$, $8\times8$,
$10\times10$, and $12\times12$ grid. The {\bf top} diagrams show the dependence
on the regularization parameter for a $5\times5$ grid. With increasing $l$ the
lines are for $\lambda=$ 100, 10, 1, 0.1, and 0.01. However, the latter 
likelihood values are close together, and so the last two or 
three lines cannot be distinguished in this representation.
[The values for ${\rm Max}(l)$ indicated by the vertical
lines were calculated using equation (\ref{papp}) for the probability 
distribution of image ellipticities, whereas the histograms were determined
without this approximation. The differences between approximation and
exact calculation are small and do not change any of the conclusions drawn in
the text.]}
\label{loli}
\end{figure}

Finally Fig.~\ref{loli} depicts the evolution of the maximum ${\rm Max}(l)$ 
of the logarithm of the likelihood function when changing the 
regularization parameter and the number of grid points. In the case of the 
$5\times5$ grid the rather large separation between the grid points already 
provides some kind of regularization to prevent the cluster component from 
fitting noise. Decreasing $\lambda$ increases the maximum of the likelihood 
function only slightly, and ${\rm Max}(l)$ stays within the
probability distribution $p(l)$ for the logarithm of the likelihood
that indicates statistical consistency with a correct description of
the mass distribution.    
Although the mass reconstructions within the range of regularization values 
tested here ($\lambda=$ 0.01 -- 100) are all statistically consistent with the
observables in an absolute likelihood sense, we consider $\lambda=1$ as the
optimal value providing the most reliable results for the analysis of our 
simulations. Regularizing too strongly causes the problems that were 
discussed above. Very small $\lambda$ values lead to an extension of the 
likelihood contours in the direction of very high values for the cut-off radius
parameter (see Fig.~\ref{like4}). Visual inspection of the reconstructed 
cluster component for these large cut-off radius solutions reveals strongly
fluctuating mass distributions which obviously fit noise.

Increasing the number of grid points allows the cluster component to adjust to
smaller structures, but it also increases the tendency to fit noise. The
dependence of ${\rm Max}(l)$ on the grid size (for fixed $\lambda$) is much
stronger than its dependence on the regularization parameter.
For a $12\times12$ grid with $\lambda=1$ the value of ${\rm Max}(l)$ is already
far in the tail of its expected distribution, which suggests that small scale
structures in the mass distribution of the reconstructed cluster component
mainly consist of noise. To compensate this effect one would need to
increase the regularization strength.
The figure also shows that in the case of extended dark matter haloes for the
galaxies, the difference in the likelihood values between the best
reconstruction with galaxies and a reconstruction without galaxies is
much larger than for the input model with small cut-off radius for the cluster
galaxies.

\section{application to Cl0939+4713}\label{appreal}
\subsection{Cluster reconstruction}
We test the method described above on a WFPC2 image (Dressler et al. 1994b) of
the cluster Cl0939+4713. From these observations Seitz et al. (1996, hereafter
SKSS) obtained a reconstruction of the cluster mass distribution by applying a
non-linear finite-field inversion algorithm. We use their galaxy catalogue
which supplies positions, ellipticities, and $R$ magnitudes. The cluster is at
a relatively high redshift of $z_d=0.41$. This means that the normalization of
the reconstructed mass distribution is rather sensitive to the unknown redshift
distribution of the faint galaxies that are used as sources. In addition, due
to the small field of view which contains the central cluster region only, it
is not possible to break the mass sheet transformation by assuming negligible
surface mass density values at the boundaries. The lack of colour
information for the objects detected in the field also makes it difficult to
break this degeneracy by using magnification effects on the number counts or on
the sizes of the background galaxies.  
SKSS investigated the implications of these problems on their cluster
reconstruction. Here we restrict the calculations to assumptions for
the relevant parameters and qualitatively discuss the effects of changing them
afterwards. The cluster is marginally critical and contains an arc as well as a
multiple image system of a source galaxy located at a redshift of 3.98 (Trager
et al. 1997). The extension of the critical region is very small compared to
the total area of the field, and so we exclude it from our analysis. 

\begin{figure*}
    \epsfxsize=150mm
    \epsffile{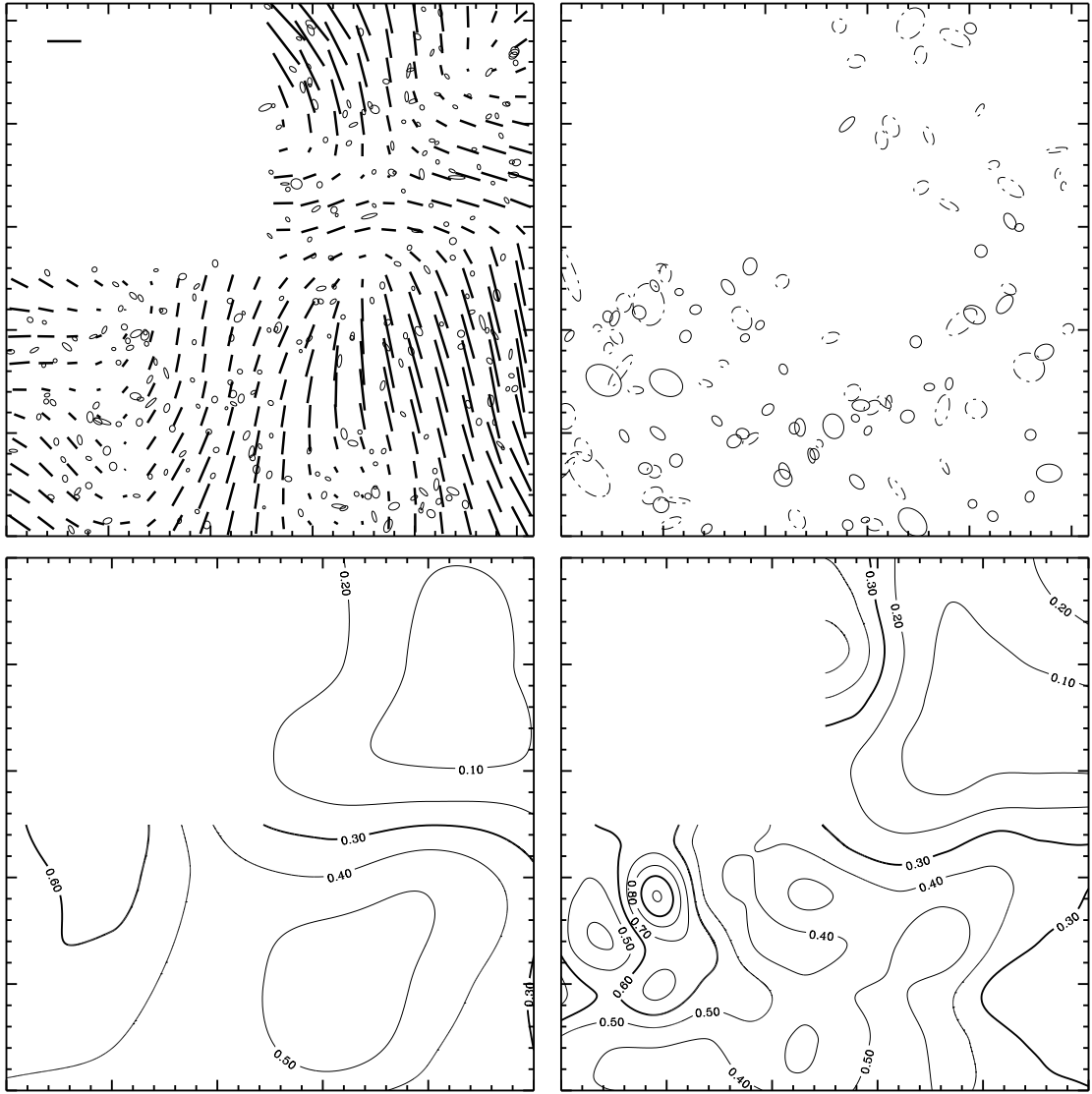}
\caption{The galaxy and mass distribution in the WFPC2 field of the cluster 
Cl0939+4713. The {\bf top left} plot displays the galaxies in the magnitude
interval $R\in[23.5,25.5]$ which were used as sources for the weak lensing
analysis. Each galaxy is represented by an ellipse which indicates its
ellipticity, position angle and relative size. Superposed on these individual 
galaxy images is their average ellipticity field on a $20\times 20$ grid.   
The length of the bar in the top left corner of this plot corresponds to an
average ellipticity of $|\overline\epsilon|=0.1$. The {\bf top right} plot
shows all the galaxies that were included as cluster galaxies in our analysis. 
The ellipses drawn with solid lines represent objects which were classified as
E- or S0-galaxies in Smail et al. (1997), and the broken lines indicate those
classified as spiral or irregular galaxies. The {\bf bottom} plots depict
reconstructions of the cluster mass distribution using our maximum likelihood
method without including the presence of cluster galaxies. The {\bf
left} plot was produced with a $6\times6$ grid, and the plot on the
{\bf right} was obtained with a $12\times12$ grid. For the purpose of 
this representation the reconstructions have been bicubically interpolated on a
finer grid before using the contour plot program. The regularization parameter
was $\lambda=10$. The contour lines depict the surface mass density values
$\kappa_\infty=0.1,0.2,0.3,\dots,1.0$. The mass sheet degeneracy was
artificially adjusted such that the average surface mass density in
the data region is $\bar\kappa_\infty=0.35$. The side length of the
field of view is $2\farcm58$ which corresponds to $505h^{-1}\,{\rm
kpc}$ at the cluster redshift of $z_d=0.41$. (We assume $\Omega=1$ and
$\Lambda=0$ in this paper.)} 
\label{cl0939}
\end{figure*}

We consider all galaxies in the magnitude interval $R\in[23.5,25.5]$ that 
fulfill the observational selection criteria of SKSS as sources. However, we
exclude three objects whose images are located between the central cluster 
galaxies and are therefore likely to lie in the critical region. In
total we use 276 galaxy images. Fig.~\ref{cl0939} shows the spatial
distribution of this population and the ellipticity field that was
calculated from their shapes by employing an averaging procedure with
a scale of $20\arcsec$. We assume a source redshift distribution of
the same form as it was used for our simulations, i.e., equation (10)
of Paper~I with the parameters  $\langle z\rangle=1$ and
$\beta=1$. This gives a value of $\langle w\rangle=0.44$ for the
average relative lensing strength (we use $\Omega=1$ and
$\Lambda=0$). In analogy to SKSS we reconstruct the cluster mass from
the ellipticity field by applying the methods developed in Seitz \&
Schneider (1996) and Seitz \& Schneider (1997). Our result is
consistent with theirs. We scale the reconstructed mass distribution
such that the average surface mass density is $\bar\kappa_\infty=0.35$
by using the mass sheet transformation in the form generalized to a
redshift distribution of the sources (Seitz \& Schneider 1997). This
gives a total mass of $1.97\times10^{14}h^{-1}\,\rm{M}_{\sun}$ within
the field of view. 

We calculate the lensing potential from the surface mass density. This involves
an integration that formally extends over the whole lens plane, whereas the
data region is limited. The shear contribution provided by the matter outside
of the field of view cannot be neglected here, because the surface mass density
values are still appreciable at the boundary. We extrapolate the reconstructed
mass distribution with a simple linear prescription outside of the
data region before calculating the potential $\psi_{\alpha\beta}$ on a
grid. Therefore, the result correctly describes the mass distribution
in the field of view as it was determined from 
the reconstruction, but it contains only an approximate description of the
shear. Nevertheless, it serves as a reasonable start value for the maximum
likelihood algorithm. The arbitrary element introduced by the extrapolation 
procedure is then of course resolved during the likelihood maximization which 
adjusts the potential such that the shear is optimally reproduced.  
We use the surface mass density values on the grid points calculated from the
initial potential grid as the regularization prior 
$\kappa_{\rm P\alpha\beta}$. The missing quadrant of the wide field camera does
not cause a problem for the technical implementation of the method. The
potential values on the irrelevant grid points are excluded from the
maximization procedure. The calculation of the entropy $S$ and the average
surface mass density  $\bar\kappa_{\rm C}$ is restricted to the data
region. For the probability distribution of image ellipticities the
approximation (\ref{papp}) is applied with dispersions calculated in analogy to
Fig.~\ref{gred}, taking into account the different cluster redshift and using
the same intrinsic ellipticity dispersion of $\sigma_{\epsilon_s}=0.2$ as for
the simulations.  

Fig.~\ref{cl0939} shows maximum likelihood reconstructions of the cluster
Cl0939+4713 that were performed with our algorithm without taking the 
presence of the cluster galaxies into account. Using a $6\times6$ grid for
the surface mass density gives a grid point separation of 
$101h^{-1}\,{\rm kpc}$ and allows to resolve the main features characterizing
the mass distribution. As discussed by SKSS the two mass maxima in the bottom
quadrants and the minimum in the top right quadrant tend to correlate with the
distribution of bright galaxies. This trend is also visible in the diagram 
depicting the distribution of our cluster galaxy sample which will be
defined below. The galaxies in the top right quadrant are mainly less bright
spirals and a substantial fraction of them might not be cluster members.
A $12\times12$ grid with a grid point separation of $46h^{-1}\,{\rm kpc}$ 
enables the mass distribution to adjust to smaller structures. The
weak lensing reconstruction now produces a high mass
peak roughly at the position of the strong lensing features. The height of this
maximum depends on the regularization strength. Its exact position depends on
the number of grid points, because in the method maxima can obviously only
occur at grid points. The positional uncertainty of the mass peak is such that
its location is consistent with coinciding with the strong lensing region.
SKSS were not able to provide evidence for such a rather steep increase of the
surface mass density at this position, because inversion algorithms which rely
on the averaged ellipticity field as input tend to smear out such features,
whereas the maximum likelihood method applied here uses the individual image
ellipticities directly as observables (see also Seitz et~al. 1998).
However, the reliability of the smaller scale structures in this mass
map is doubtful and their significance is difficult to estimate. We
will come back to this question at the end of the next section in
context with a discussion of the absolute likelihood value. 

The strong lensing features are caused by the joint effects of a
global cluster component and several bright cluster galaxies located
in the cluster centre. We do not attempt to reproduce the multiple
image system here, because the spatial resolution reliably accessible to the
non-parametric reconstruction is considerably larger than the image
separation. This also means that it is very difficult to use the
strong lensing information for breaking the mass sheet degeneracy and
to determine the slope of the central cluster profile in this case. 

\subsection{Cluster galaxy analysis}\label{clga}
To select cluster galaxies we used the morphological classification and
instrumental magnitudes ($R_{702}$) provided by Smail et al. (1997). 
Galaxies that were denoted as `compact' or `unclassifiable', or for which 
measured redshifts from Dressler \& Gunn (1992) exclude a cluster
membership, were not taken into account. We define an $L_{\star}$-galaxy 
by requiring an absolute V-magnitude of $-20.5+5\log h$ 
(corresponding to $m_{V}=20.2$), which is close to the value
determined by Smail et al. (1997) from a fit to the luminosity
function of elliptical galaxies in clusters of  
this redshift range. For direct comparison we take the same value for spiral 
galaxies as well. For the cluster redshift of
$z_d=0.41$ the R-band filter probes roughly the same spectral range as
a rest frame V-band filter, and the correction factors for the
conversion of these filters are small (B. Ziegler, private
communication). For our cluster galaxy sample we therefore 
identify R-magnitudes from the catalogue with rest-frame V-magnitudes. 
Possible systematic errors introduced by this simplification are in any
case smaller than those resulting from other problems that will be
discussed later. As in our simulations we only include galaxies with a 
luminosity brighter than $0.1L_{\star}$ ($m_{V}<22.7$) in our
analysis, because very faint cluster galaxies do not contribute much
to the lensing signal. 
According to the morphological classification we divide our sample into two 
subsets. One of them includes 56 elliptical and S0-galaxies and the other 
contains 55 objects classified as spiral or irregular galaxies. 
Fig.~\ref{cl0939} schematically displays the galaxies that were
included in our cluster galaxy sample. It is very likely that some of
the objects in the sample are in fact fore- or background galaxies
and the implications of this problem on the results of the lensing
analysis will be discussed below. Fig.~4 of Dressler et al. (1994a)
and Fig.~6 of Belloni et al. (1995) suggest that especially in the
spiral galaxy subset a significant fraction could be interloping field
galaxies.  

\begin{figure}
    \epsfxsize=84mm
    \epsffile{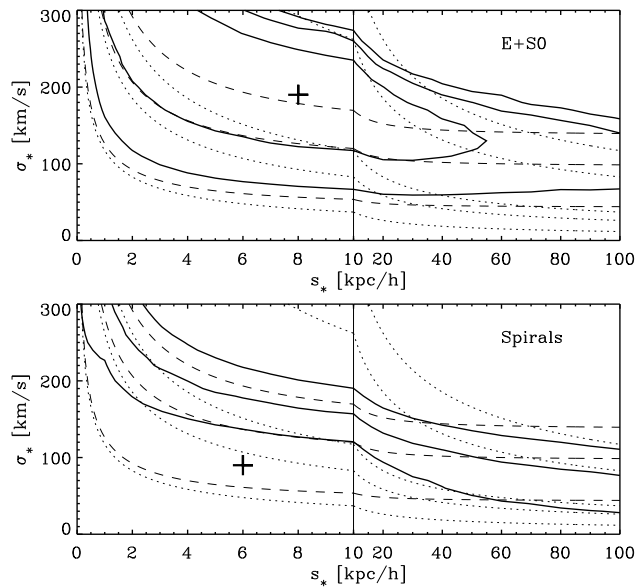}
\caption{Results of applying the galaxy-galaxy lensing method presented in this
paper to a WFPC2 image of the cluster Cl0939+4713. The {\bf top} plot shows
confidence contours for the parameters describing the halo properties of an
ensemble of 56 elliptical cluster galaxies, and the {\bf bottom} plot is for
55 spiral galaxies. Note the change of scale at 
$s_{\star}=10h^{-1}\,{\rm kpc}$. The calculations were performed with a
$6\times6$ grid for the cluster mass distribution, a regularization parameter
of $\lambda=100$, and the values $\eta=4$ and $\nu=0.5$ for the parameters of 
the scaling laws relating the properties of galaxies with different
luminosities to those of an $L_{\star}$-galaxies, which we defined as
having an absolute V-magnitude of $-20.5+5\log h$. The likelihood
function includes shape information from 276 background galaxy
images. The confidence contours 
are 99.73\%, 95.4\%, and 68.3\%, determined in the way explained in Paper~I,
and the cross marks the maximum of the likelihood function. The dashed lines
connect models with an equal aperture mass of $M_{\star}(<8h^{-1}\,{\rm
kpc})=0.1,0.5,$ and $1.0\times 10^{11}h^{-1}\,{\rm M}_{\sun}$,
respectively. Similarly, the dotted lines are for a total $L_{\star}$-galaxy
mass of $M_{\star}=0.1,0.5,1.0,5.0$ and $10\times 10^{11}h^{-1}\,{\rm
M}_{\sun}$, which corresponds to a `galaxy mass fraction' of
$0.15\%,0.75\%,1.5\%,7.5\%$ and $15\%$, respectively. {\it As discussed in
the text, the results displayed here depend on assumptions regarding
the redshift distribution of the sources and the mass sheet
transformation! Changing these assumptions mainly leads to a shift of
the confidence contours in $\sigma_{\star}$-direction.}}  
\label{result}
\end{figure}

Fig.~\ref{result} displays the results of applying the galaxy-galaxy lensing
method developed in Section~\ref{meth}. Here we chose a $6\times6$ grid for 
the cluster mass distribution which provides sufficient resolution
for an adequate description of the reliably reconstructed features.   
As a start value for the potential $\psi_{\alpha\beta}$ we took the maximum
likelihood solution without galaxies that has been discussed in the
previous section. In addition, the corresponding cluster mass
distribution was taken as the regularization prior. The regularization
parameter was chosen as $\lambda=100$. In analogy to our simulations
we use the scaling laws (\ref{scaling})
to relate the velocity dispersion and the cut-off radius of galaxies
with different luminosity to those of an $L_{\star}$-galaxy, and we
fix the scaling indices at the values $\eta=4$ and $\nu=0.5$.
We performed the calculation for the ensemble of 56 elliptical galaxies
ignoring the presence of the spiral galaxies at first. For given values of the
galaxy model parameters the method searches for the best representation of the
remaining cluster mass component leaving the total mass of the system constant.
Reassuringly, the resulting likelihood contours for the galaxy model
parameters $\sigma_{\star}$ and $s_{\star}$ show qualitatively the same
properties as in our simulations. Despite the small number of lens and source
galaxies, the lensing effects of the elliptical cluster galaxies are detected
with high significance: Reducing the mass of the cluster component and
putting it into galaxies increases the likelihood of the observed
image ellipticities of background galaxies. As expected, the
likelihood contours are extended along lines connecting models that
imply equal mass for an $L_{\star}$-galaxy within some characteristic
radius. We empirically determined a value of $8h^{-1}\,{\rm kpc}$ by
adjusting the dashed lines in the figure to the shape of the
confidence contours in the low-$s_{\star}$ region of the
plot. Therefore the mass within this radius is the best-determined  
quantity from this analysis. 

The figure also includes the result of an equivalent analysis of the
spiral galaxy sample. In contrast to the ellipticals we do not find a
signal in this case, although the number of galaxies and
the range of luminosities are comparable for the two sets of galaxies. 
However, there is a factor of $\sqrt 2$ difference between the
characteristic velocity dispersion parameter for elliptical and spiral
galaxies of the same luminosity. This translates into a factor of two 
difference in the expected strength of the lensing effects, and so it
is not surprising that the result is weaker for the spiral galaxy sample. 
As was mentioned above some of the spirals in this sample might actually be
background galaxies, whereas we are interested in the properties of
cluster galaxies here. Of course this could lead to a bias of the
result, simply because they are the wrong kind of objects included in
the sample, but also because their size and luminosity, as well as the
strength of their lensing effects on the population of faint source
galaxies would then be different than tested by the model, which
assumes that they are located at the cluster redshift.  
In contrast to that, not taking into account cluster galaxies, which
actually are present, does not introduce systematic effects: Repeating
the above calculation of the confidence contours for spirals and ellipticals
with inclusion of the best-fit model for the other galaxy type only leads
to negligible changes of the result. The fact that non-cluster
galaxies that were misclassified as cluster members can cause systematic
effects whereas leaving out genuine cluster members does not, suggests
that it is useful to apply conservative selection criteria for
defining the cluster galaxy sample. 

\begin{figure}
    \epsfxsize=84mm
    \epsffile{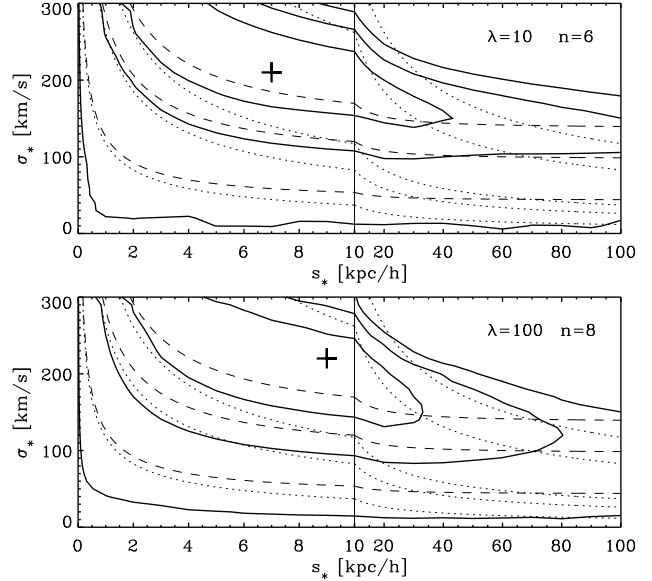}
\caption{Dependence of the result for elliptical cluster galaxies on the number
of grid points and the regularization strength. The {\bf top} plot is for
$\lambda=10$ with a $6\times6$ grid, and the {\bf bottom} one for $\lambda=100$
with an $8\times8$ grid. The significance of the contours and the
meaning of the lines is analogous to the previous figure.}
\label{check}
\end{figure}

We checked the sensitivity of the result to changes of the regularization 
strength or the number of grid points that are used to describe the
cluster potential. Fig.~\ref{check} displays the confidence contours
for a reduced regularization parameter of $\lambda=10$ (keeping a
$6\times6$ grid), as well as for an $8\times8$ grid for the cluster
component (keeping $\lambda=100$). These calculations necessitate a
larger amount of computation time than those performed for Fig.~\ref{result}.
The appearance of the confidence regions does not change very much
with respect to our previous result. Now the axes of the plots are
outside of the $99.73\%$ confidence contour in both cases. (The axes
correspond to model parameters implying zero mass in galaxies and
therefore represent a pure cluster mass reconstruction without cluster
galaxies.) This means that the galaxy-galaxy lensing signal becomes
slightly more significant when the ability of the cluster component to
adjust to the  presence of the cluster galaxies is increased either by
improving the resolution or by reducing the regularization
strength. The confidence contours in Fig.~\ref{check} (but not those
in Fig.~\ref{result}) have been slightly smoothed in order to erase
some discontinuities caused by a premature stopping of the
maximization algorithm for some parameter values. This becomes a more
serious problem for a larger number of grid points or a smaller
regularization parameter.   
For the refined resolution achieved with the $8\times8$ grid, which
corresponds to a grid point separation of $72h^{-1}\,{\rm kpc}$, the
confidence contours are less extended along the cut-off radius coordinate.
As we discussed in Section~\ref{appsim} this effect can be explained
by the ability of the cluster component to adapt to mass structures on
scales comparable to the largest cut-off radius values shown in the plot.

To perform our analysis we were obliged to make assumptions on the redshift 
distribution of the faint objects in our background galaxy sample.
Underestimating the source redshifts leads to an overestimate of the
lens masses and therefore to a shift of the cluster galaxy confidence
contours towards higher values for the velocity dispersion
parameter. Conversely, overestimating the source redshifts leads to an
underestimate of the galaxy masses. However, in our application the mass sheet
degeneracy constitutes an additional complication. We broke this
degeneracy artificially by fixing the average surface mass density at
the value $\bar\kappa_{\infty}=0.35$. For the redshift distribution that
we have assumed this is roughly the minimal value that ensures that
the mass density is non-negative over the whole field of view.
Changing the assumptions for the redshift distribution leads to a
change of the minimal possible value for the source redshift-independent
quantity $\bar\kappa_{\infty}$.
Equally viable solutions with larger $\bar\kappa_{\infty}$ provide more 
convergence and require smaller variations in the mass distribution --
and thus smaller galaxy masses -- to produce the same observable image
distortions. Adjusting $\bar\kappa_{\infty}$ therefore allows to scale
the resulting confidence contours for the cluster galaxy model
parameters in the direction of the $\sigma_{\star}$-axis of the plot.
In future applications with larger fields of view, dealing with this
technical problem will hopefully become obsolete.
The problems mentioned here make it impossible to obtain tighter
limits on the cut-off parameter $s_{\star}$ by including prior
knowledge on the velocity dispersion parameter $\sigma_{\star}$ as it
was envisaged in Paper~I.

The procedure that had been used by SKSS to determine image
ellipticities for the faint objects in the source galaxy catalogue
did not include a detailed analysis of instrumental properties as it was
carried out by Hoekstra et~al. (1997). However, typical distortion
values in the field we analysed are considerably larger than their instrumental
corrections and so we believe that the implications on our results are
negligible. 

\begin{figure}
\begin{center}
    \epsfxsize=50mm
    \epsffile{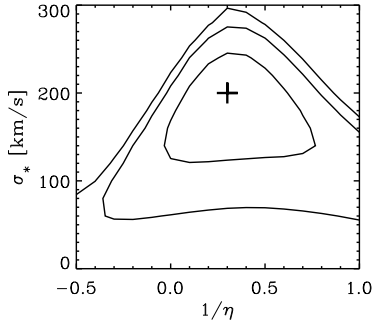}
\end{center}
\caption{Constraints on the scaling parameter $\eta$ for elliptical cluster
galaxies. The plot shows 99.73\%, 95.4\%, and 68.3\% confidence contours for
the velocity dispersion $\sigma_{\star}$ and the scaling index $\eta$. 
Here we assumed a characteristic cut-off radius of
$s_{\star}=8h^{-1}\,{\rm kpc}$, corresponding to the maximum of the  
likelihood function in Fig.~\ref{result}, and a cut-off radius scaling index of
$\nu=0.5$. The number of grid points is $6\times6$ and the regularization
parameter is $\lambda=100$.}
\label{scale}
\end{figure}

To obtain the results shown so far we chose $\eta=4$ for the index of
the scaling law that relates the velocity dispersion parameter of
galaxies with different luminosities to those of an $L_{\star}$-galaxy.  
This choice was motivated by the observed Faber-Jackson relation for the
velocity dispersion of the stars in elliptical galaxies. If we assume a
roughly isothermal mass density profile for the galaxies a similar
relation should also hold for the velocity 
dispersion parameter $\sigma$ of their dark matter halo. In Fig.~\ref{scale}
we explored the sensitivity of the result to the choice of the scaling
index. The figure shows confidence contours in the parameter space
defined by the velocity dispersion $\sigma_{\star}$ and the scaling
index $\eta$. The value of $\eta=4$ lies within the $68\%$ confidence 
contour and the likelihood difference compared to the maximum of the
likelihood function is insignificant. We consider this figure as a
necessary check if the information provided by lensing is consistent
with our observationally motivated assumptions, and not as a serious
attempt to constrain the parameter $\eta$ from the scarce data available here. 
For this plot the value for the cut-off radius parameter
was fixed at $s_{\star}=8h^{-1}\,{\rm kpc}$, which corresponds to the
maximum of the likelihood function in Fig.~\ref{result}. Changing
$s_{\star}$ favours different values for the velocity dispersion parameter
and leads to a shift of the contours along the $\sigma_{\star}$-coordinate 
in Fig.~\ref{scale}. However, this does not affect the conclusions
regarding the scaling index that we made above.
Another parameter whose importance for our results we did not yet
discuss is the index $\nu$ for the cut-off radius scaling law. In
contrast to the parameter $\eta$ we do not have a sound motivation for
a particular choice of this index. Changing this parameter between
$\nu=0$ and $\nu=1$ only leads to marginal differences in the
likelihood and so it is impossible to achieve limits on this parameter 
from these data. In addition, the implications on the appearance of
the confidence contours in Figs.~\ref{result} to \ref{scale}, for
which $\nu=0.5$ was assumed, are very small.

We now turn to a discussion of the absolute likelihood level of our
reconstructed solutions. In order to determine the range of values for
$l$ that indicate statistical consistency, we generate random
intrinsic ellipticities for the objects in our source galaxy catalogue. 
Taking the reconstructed solution for the cluster we then calculate
their lensed image ellipticities and from that the value of the
logarithm of the likelihood. Repeating that procedure for many
different intrinsic realizations allows to construct a histogram for
$p(l)$ corresponding to those shown in Fig.~\ref{loli} for the simulations.
As a result we obtain $l$-values approximately between 260 and 300.
However, the $6\times6$ grid reconstruction without galaxies presented in
Fig.~\ref{cl0939} only reaches a log-likelihood level of $l=138$
for the actually observed image ellipticities, which is far below the
value tolerated for an acceptable solution. 
Including cluster galaxies allows to increase this value by 5 or 6, but
refining the cluster grid is much more effective in this respect. 
The $12\times12$ grid solution also shown in Fig.~\ref{cl0939} attains
a value as high as $l=172$, which however still falls short of the expected
level by a large amount. (The expected level as determined above does
not depend very much on the choice for the number of grid points.)
In theory the number of grid points and the regularization strength
should be adjusted such that consistent likelihood values are reached.
This works quite well for our application to simulated observations
described in Section~\ref{appsim} where we assumed that the
probability density distributions for the intrinsic ellipticity and
the redshift of the source galaxies are known.
For the application to real data discussed here, however, this is not
the case and probably the problem that we face here -- and that is reflected
by the discrepant log-likelihood levels -- is not an inadequate description
of the cluster mass distribution, but rather an insufficient knowledge
of the properties of the faint galaxies that we have used as lensed
sources for our study. In particular we made simple assumptions
regarding the redshift distribution and the intrinsic ellipticity
distribution of these galaxies. In practice these could be quite
different for galaxies of different magnitude, surface brightness or
morphology. In order to use a check of the absolute likelihood level
as a powerful tool for evaluating the reliability of our description
of the cluster mass distribution, a detailed understanding of the
source galaxy population is warranted. 

\section{discussion}\label{disc}
Continuing our discussion of Paper~I we investigated methods to
constrain the mass distribution of cluster galaxies by weak lensing.
In Paper~I we concluded that a comprehensive method should
simultaneously take into account the lensing effects of individual
cluster galaxies as well as those of a general cluster component.
To this end we developed a regularized cluster lens reconstruction
algorithm that directly uses the image ellipticities of individual
background galaxies as observables, and that allows to explicitly
include the presence of cluster galaxies in the analysis. 
We parametrize the mass distribution of cluster galaxies by truncated
isothermal spheres and apply simple scaling laws to relate the halo
properties to those of $L_{\star}$-galaxies according to the galaxy
luminosity. For each value of the galaxy parameters -- velocity
dispersion $\sigma_{\star}$ and cut-off radius $s_{\star}$ of an
$L_{\star}$-galaxy -- the method allows to determine the best
representation for the underlying cluster mass component that is 
consistent with the observed image ellipticities. We tested the method
on simulations and achieved convincing and robust results.
If the dark matter distribution around cluster galaxies is rather extended, 
it is not straightforward to make a clear-cut distinction between mass 
belonging to a global cluster component or to galaxy haloes.  
An important result from these simulations is that the distance between the
grid points of the cluster component determines the scale for a
somewhat artificial separation of the mass located in galaxies and the
mass attributed to the global cluster component.

We applied our method to a WFPC2 image of the galaxy cluster Cl0939+4713.
Our reconstruction of the cluster mass distribution is consistent with
the result of SKSS who used the same object catalogue but a different 
reconstruction technique. However, the robustness of the result should
finally be verified by using observations with a larger field of view. This
would also allow to break the mass sheet degeneracy that prevents us
from deriving absolute mass estimates here. In this respect another
factor of uncertainty is the unknown redshift distribution of the
source galaxies.

The galaxy-galaxy lensing analysis provided a significant detection 
(approximately at the 99\% to 99.9\% confidence level) of the lensing 
effects induced by luminous elliptical cluster galaxies. In fact it
had already been speculated by Dressler et~al. (1994b) that some of
the arclets in this WFPC2 image `appear to be background sources
lensed by individual galaxies rather than the overall cluster potential'.
The likelihood function attains its maximum at rather low values for
the cut-off radius parameter. However, the small number of lens galaxies 
included in our study does not allow to derive strong limits for the
parameters describing the galaxy haloes. Changing the assumptions that
we were obliged to make regarding the mass sheet degeneracy and the
redshift distribution of the sources leads to a rescaling of the
galaxy-galaxy lensing result. For this reason we do not quote galaxy
mass-to-light ratios or values for the fraction of the total mass
bound to galaxies in this paper. Our result is consistent with the
result of Natarajan et~al. (1997) who applied the methods of Natarajan
\& Kneib (1997) to analyse a mosaic of WFPC2 images of the cluster
AC114. However, significant differences of their approach compared to ours
make a quantitative comparison of the results difficult. 

We consider the application to Cl0939+4713 described in this paper as
a successful test of our maximum likelihood technique for galaxy-galaxy
lensing in clusters of galaxies. The problems we encountered are not
inherent to our method and mainly resulted from the small size
of the image we analysed. Ideal observations for this project
should include deep exposures with high image quality and large fields
of view, which would make it feasible to test possible variations of
the halo properties with the density of the environment. Another
important ingredient should be supplementary imaging in different colours in
order to assure a reliable selection of cluster galaxies and
potentially for deriving photometric redshift estimates for the
galaxies used as sources. 
The benefit of photometric redshift information for galaxy-galaxy
lensing studies (in the field) was emphasized by Schneider \& Rix
(1997) and has been demonstrated by Hudson et~al. (1997) in their
analysis of the {\it Hubble Deep Field}\/. Making use of available redshift 
information for individual source galaxies also allows us to reduce
the noise in cluster reconstructions. [In our implementation the approximation
(\ref{papp}) for the probability density distribution of image ellipticities
can then be replaced by the simpler expression (\ref{pappsimp}).]
This was also recognized by Seljak (1997) in a similar context.

The analysis of the kind of observational data sets envisaged above,
which are most likely to be available within the next few years, will
provide the opportunity to compare the lensing constraints to recent
numerical studies of the properties of galaxy haloes within clusters 
(Tormen, Diaferio \& Syer 1997, Ghigna et~al. 1998). In addition, a comparison
and combination of the galaxy-galaxy lensing results with morphological 
studies of the galaxy population in these clusters (e.g. Oemler,
Dressler \& Butcher 1997) could help to obtain observational evidence
for the physical processes [such as the `galaxy harassment' picture
suggested by Moore et~al. (1996)] that might be responsible for the
evolution of cluster galaxies.   

\section*{acknowledgments}
We thank Simon White for carefully reading the manuscript. This work was
supported in part by the Deutscher Akademischer Austauschdienst
(Doktorandenstipendium HSP III), the Sonderforschungsbereich 375-95
der Deutschen Forschungsgemeinschaft and the TMR Network on Research
in Gravitational Lensing.

\appendix

\section{}\label{appprob}
In this appendix we prove a statement given in Section~\ref{spdd}
concerning the dispersion of the probability density distribution for
the image ellipticity of lensed background galaxies.
Seitz \&~Schneider (1997) showed that the moments of the complex ellipticity
parameter $\epsilon$ can be expressed in terms of the reduced shear:
\begin{equation}\label{exg}
\langle\epsilon^n\rangle_{\epsilon_{\rm s}}=g^n\ \ \ {\rm for}\ \ \ |g|\leq1\;.
\end{equation}
In the following we perform a rotation of the ellipticity coordinates
such that the imaginary component of the (reduced) shear vanishes, i.e., we
transform $g$ and $\epsilon$ into `local shear coordinates'. This can be 
expressed as 
\[
\tilde g:=\frac{\gamma^{\star}}{|\gamma|}\,g=\frac{|\gamma|}{1-\kappa}
\ \ \ \ \mbox{and}\ \ \ \
\teps=\teps_1+{\rm i}\,\teps_2:=\frac{\gamma^{\star}}{|\gamma|}\,\epsilon\;.
\]
We calculate the expectation value of 
$\teps^2=\teps_1^2-\teps_2^2+2\teps_1\teps_2\,{\rm i}$ 
by applying equation (\ref{exg}):
\begin{equation}\label{exg1}
\langle\teps^2\rangle_{\epsilon_{\rm s}}=
\langle\teps_1^2\rangle_{\epsilon_{\rm s}}-
\langle\teps_2^2\rangle_{\epsilon_{\rm s}}+
2\langle\teps_1\,\teps_2\rangle_{\epsilon_{\rm s}}\,{\rm i}=\tilde g^2\;.
\end{equation}
The quantity $\tilde g$ is real and so it follows from this equation that
$\langle\teps_1\,\teps_2\rangle_{\epsilon_{\rm s}}=0$. Now we can express the
variance of $\tilde\epsilon_1$ as
\[
\sigx^2:=
\langle\teps_1^2\rangle_{\epsilon_{\rm s}}-
\langle\teps_1\rangle^2_{\epsilon_{\rm s}}=
\langle\teps_1^2\rangle_{\epsilon_{\rm s}}-\tilde g^2
=\sigy^2\;.
\]
In the first equality we used equation (\ref{exg}) and the second follows from
equation (\ref{exg1}) with the definition of 
$\sigy^2:=\langle\teps_2^2\rangle_{\epsilon_{\rm s}}-
\langle\teps_2\rangle^2_{\epsilon_{\rm s}}$ and 
$\langle\teps_2\rangle_{\epsilon_{\rm s}}=0$ which again follows from
(\ref{exg}). Thus the dispersion of the probability density distribution in
local shear direction is equal to the dispersion in the perpendicular
direction: $\sigx=\sigy$. For $|g|>1$ this result can be derived analogously.

\section{}\label{deriv}
This appendix gives the expressions needed to calculate the derivatives
of $\hat l$ with respect to the potential values on the grid points. The 
derivative of the logarithm of the probability density can be written as
\begin{eqnarray}
\frac{\partial\ln p_{\epsilon}(\epsilon)}{\partial\psi_{\alpha\beta}} & = &
\frac{\partial\ln p_{\epsilon}(\epsilon)}{\partial\kappa_{\infty}}
\frac{\partial\kappa_{\infty}}{\partial\psi_{\alpha\beta}}  \nonumber \\
& + & 
\left[\frac{\partial\ln p_{\epsilon}(\epsilon)}{\partial|\gamma_{\infty}|}
\frac{\gamma_{1\infty}}{|\gamma_{\infty}|}-
\frac{\partial\ln p_{\epsilon}(\epsilon)}{\partial\phi_{\gamma}}
\frac{\gamma_{2\infty}}{|\gamma_{\infty}|^2}
\right]\frac{\partial\gamma_{1\infty}}{\partial\psi_{\alpha\beta}}  
\nonumber \\ & + &
\left[\frac{\partial\ln p_{\epsilon}(\epsilon)}{\partial|\gamma_{\infty}|}
\frac{\gamma_{2\infty}}{|\gamma_{\infty}|}+
\frac{\partial\ln p_{\epsilon}(\epsilon)}{\partial\phi_{\gamma}}
\frac{\gamma_{1\infty}}{|\gamma_{\infty}|^2}
\right]\frac{\partial\gamma_{2\infty}}{\partial\psi_{\alpha\beta}}\;. \nonumber
\end{eqnarray}
The symbol $\phi_{\gamma}$ represents the phase of the complex shear parameter 
$\gamma_{\infty}=|\gamma_{\infty}|\,{\rm e}^{{\rm i}\,\phi_{\gamma}}$. The 
derivative of $\ln p_{\epsilon}(\epsilon)$ with respect to $\kappa_{\infty}$ 
is given by
\begin{eqnarray}
\frac{\partial\ln p_{\epsilon}(\epsilon)}{\partial\kappa_{\infty}} & = &
\frac{\teps_1-\gred}{\sigx^3(\gred)}\left[
\frac{\partial\gred}{\partial\kappa_{\infty}}\sigx(\gred)+(\teps_1-\gred)
\frac{\partial\sigx(\gred)}{\partial\kappa_{\infty}}\right]  \nonumber \\
& + &
\frac{\teps_2^2}{\sigy^3(\gred)}
\frac{\partial\sigy(\gred)}{\partial\kappa_{\infty}}
\nonumber \\ & - &
\frac{1}{\sigx(\gred)}\frac{\partial\sigx(\gred)}{\partial\kappa_{\infty}}-
\frac{1}{\sigy(\gred)}\frac{\partial\sigy(\gred)}{\partial\kappa_{\infty}}\;, 
\nonumber
\end{eqnarray}
a similar expression holds for 
$\partial\ln p_{\epsilon}(\epsilon)/\partial|\gamma_{\infty}|$, and the 
derivative with respect to $\phi_{\gamma}$ is
\[
\frac{\partial\ln p_{\epsilon}(\epsilon)}{\partial\phi_{\gamma}}=
-\frac{(\teps_1-\gred)\,\teps_2}{\sigma_{\teps_1}^2}
+\frac{\teps_2\,\teps_1}{\sigma_{\teps_2}^2}\;.
\]
The derivatives remaining in these formulae are trivial to calculate from the
definitions of the respective quantities.
The derivative of the term that ensures the correct total mass reads
\begin{eqnarray}
\frac{\partial\ln p_{\bar\kappa_{\rm C}}}{\partial\psi_{\alpha\beta}} & = &
\sum_{\gamma,\delta=1}^{n}
\frac{\partial\ln p_{\bar\kappa_{\rm C}}}{\partial\kappa_{{\rm C}\gamma\delta}}
\frac{\partial\kappa_{{\rm C}\gamma\delta}}{\partial\psi_{\alpha\beta}}
\nonumber \\ & = &
-\frac{\bar\kappa_{\rm C}-f_{\rm C}\,\bar\kappa_{\infty}}
{\sigma_{\kappa_{\rm C}}^2}
\,\frac{1}{n^2}\sum_{\gamma,\delta=1}^{n}
\frac{\partial\kappa_{{\rm C}\gamma\delta}}{\partial\psi_{\alpha\beta}}\;,
\nonumber
\end{eqnarray}
and the entropy term yields
\[
\frac{\partial S}{\partial\psi_{\alpha\beta}}=
\sum_{\gamma,\delta=1}^{n}
\frac{\partial S}{\partial\kappa_{{\rm C}\gamma\delta}}
\frac{\partial\kappa_{{\rm C}\gamma\delta}}{\partial\psi_{\alpha\beta}}=
\sum_{\gamma,\delta=1}^{n}\sum_{\epsilon,\zeta=1}^{n}
\frac{\partial S}{\partial\hat\kappa_{{\rm C}\epsilon\zeta}}
\frac{\partial\hat\kappa_{{\rm C}\epsilon\zeta}}
{\partial\kappa_{{\rm C}\gamma\delta}}
\frac{\partial\kappa_{{\rm C}\gamma\delta}}
{\partial\psi_{\alpha\beta}}
\]
with
\[
\frac{\partial\hat\kappa_{{\rm C}\epsilon\zeta}}
{\partial\kappa_{{\rm C}\gamma\delta}}=
\frac{1}{\kappa_{{\rm P}\epsilon\zeta}}
\left(\frac{\delta_{\gamma\epsilon}\delta_{\delta\zeta}-
\hat\kappa_{{\rm C}\epsilon\zeta}}
{\sum_{\eta,\theta=1}^{n}\tilde\kappa_{{\rm C}\eta\theta}}\right)
\ \ \ {\rm and} \ \ \
\frac{\partial S}{\partial\hat\kappa_{{\rm C}\epsilon\zeta}}=
\frac{1}{\hat\kappa_{{\rm C}\epsilon\zeta}}\;.
\]
Although these expressions look rather complicated they can be conveniently
implemented technically and do not cause major computational problems.

\label{lastpage}

\end{document}